\documentclass{article}

\usepackage{setspace}
\usepackage{PRIMEarxiv}
\usepackage{amsmath,centernot}
\usepackage{amssymb}
\usepackage{dsfont}
\usepackage{subcaption}

\usepackage[utf8]{inputenc} 
\usepackage[T1]{fontenc}    
\usepackage{hyperref}       
\usepackage{url}            
\usepackage{booktabs}       
\usepackage{amsfonts}       
\usepackage{nicefrac}       
\usepackage{microtype}      
\usepackage{lipsum}
\usepackage{fancyhdr}       
\usepackage{graphicx}       
\usepackage{bbm}
\usepackage{algorithm}
\usepackage{algpseudocode}
\usepackage{longtable}
\usepackage{multirow}

\usepackage{pifont}

\usepackage{mathtools}
\DeclarePairedDelimiter\abs{\lvert}{\rvert}%
\makeatletter
\let\oldabs\abs
\def\abs{\@ifstar{\oldabs}{\oldabs*}}

\makeatletter
\newcommand*{\rom}[1]{\expandafter\@slowromancap\romannumeral #1@}
\DeclareMathOperator*{\argmax}{arg\,max}
\DeclareMathOperator*{\argmin}{arg\,min}

\title{Making accurate and interpretable treatment decisions for binary outcomes}

\author{
  Lingjie Shen \\
  Department of Methodology and Statistics \\
  Tilburg University \\
  \texttt{L.Shen@tilburguniversity.edu} \\
  \And
  Gijs Geleijnse\\
  IKNL (Integraal Kankercentrum Nederland) \\
  \texttt{g.geleijnse@iknl.nl} \\
  \AND
  Maurits Kaptein \\
  Jheronimus Academy of Data Science \\
  \texttt{m.c.kaptein@tue.nl}
}

\begin{document}

\maketitle

\begin{abstract}
Optimal treatment rules can improve health outcomes on average by assigning a treatment associated with the most desirable outcome to each individual. Due to an unknown data generation mechanism, it is appealing to use flexible models to estimate these rules. However, such models often lead to complex and uninterpretable rules. In this article, we introduce an approach aimed at estimating optimal treatment rules that have higher accuracy, higher value, and lower loss from the same simple model family. We use a flexible model to estimate the optimal treatment rules and a simple model to derive interpretable treatment rules. We provide an extensible definition of interpretability and present a method that - given a class of simple models - can be used to select a preferred model. We conduct a simulation study to evaluate the performance of our approach compared to treatment rules obtained by fitting the same simple model directly to observed data. The results show that our approach has lower average loss, higher average outcome, and greater power in identifying individuals who can benefit from the treatment. We apply our approach to derive treatment rules of adjuvant chemotherapy in colon cancer patients using cancer registry data. The results show that our approach has the potential to improve treatment decisions.
\end{abstract}


\section{Introduction}
Patients with distinct characteristics often show heterogeneity in responses to a treatment. In some cases, only a specific subgroup within a population can benefit from a treatment. Consequently, the so-called "one-size-fits-all" treatment rule doesn't apply to all individuals, and optimal treatment rules that assign a treatment with the most desirable outcome for each individual are needed \cite{collins2015new}. Currently, advanced statistical methods for handling high-dimensional data and complex interactions between patients characteristics and treatments, as well as the collection of large amount of medical data can empower estimation of the optimal treatment rules, potentially improving outcomes and lowering costs. 

The optimal treatment rules are decision rules mapping individual characteristics onto a set of treatment options that lead to highest average outcome \cite{qian2011performance}. Making the optimal treatment decision needs estimation of outcomes under each treatment (i.e., potential outcomes \cite{imbens2015causal}). There is a considerable amount of literature focusing on estimating optimal treatment rules \cite{qian2011performance,lipkovich2017tutorial,kosorok2019precision,xu2020optimal,logan2019decision,klausch2018estimating,foster2011subgroup,zhang2012estimating,zhang2012robust,zhao2012estimating,zhou2023offline,hitsch2018heterogeneous}. These methods find the optimal treatment rules that optimize an estimator of the population average outcome. These estimators include but not limited to G-computation estimators, inverse-propensity-score weighting estimators, and doubly robust estimators, etc., each of which combined with various modelling techniques. Chatton et. al. (2020) suggest that the G-computation estimator, which models conditional mean of the outcome given a treatment and covariates, is the most accurate method for estimating potential outcomes and the optimal treatment rules \cite{chatton2020g}. Furthermore, Qian and Murphy (2011) showed that good prediction accuracy of an outcome model is sufficient to ensure good performance of the associated treatment rules \cite{qian2011performance}. 

However, without prior knowledge on how the outcome is generated, constructing an outcome model is challenging. On one hand, according to our previous studies \cite{shen2020estimating, shen2023RCTrep} and related literature \cite{dorie2019automated,logan2019decision}, flexible models are more accurate in predicting the optimal treatment than simple models in both simulation and real-world cases. However, the treatment rules derived from these models are too complex, hindering them to be used in a real-life setting. On the other hand, simple models, such as linear regression, logistic regression, etc., are readily interpretable, however, these models are not flexible to capture potential nonlinearities and interactions between treatments and covariates. This limitation leads to uncorrectable bias and thus suboptimal treatment decisions \cite{hastie2009elements}. The bias can only be reduced by expanding the model space \cite{hastie2009elements}, as illustrated in Figure \ref{fig: motivation} (a) in Appendix A showing the impact of model space on bias. Motivated by this  challenge, we seek models that are both accurate and interpretable. 

Although there is a broad literature on model selection that considers both accuracy and interpretability \cite{woody2020model, peltola2018local, piironen2018projective, namkoong2020distilled,lundberg2020local,moncada2021explainable}, these studies focus on explaining machine learning models for prediction problem, while studies related to causal inference, decision making, and their application in personalized medicine to build the evidence base needed to guide clinical practice are insufficient. Therefore, this article aims to propose an approach that derives both accurate \textit{and} intepretable treatment rules and demonstrate this approach in a real clinical decision-making context. We present the general idea of our approach in Figure \ref{fig: motivation} (b) in Appendix A, which shows the potential improvement in prediction accuracy by creating a model using fitted values from an accurate model compared to creating the same model using potentially noisy observations. Our approach offers a practical insights into model selection for decision making. Instead of requiring a model fitted to a finite sample to achieve both accurate and interpretable outcomes, our approach suggests seperating these two goals, with each using the most appropriate modelling technique.  

In this study, we propose a general framework that allows flexibility in the choice of modelling techniques and loss values in treatment decisions. We demonstrate a procedure for deriving optimal treatment rules and another for simplifying these rules. We specify a flexible model to produce a good fit for potential outcomes. Based on this model and given an individual's covariates values, we assign a treatment for the individual according to which treatment yields the lowest loss. The optimal treatment rules for each individual are then derived. Then, we define a simple model family that is deemed interpretable and fit the model to the optimal treatment rule as a function of covariates. The interpretable treatment rules are then derived. In the simulation study, we demonstrate that compared to treatment rules derived from the same simple model family fitted to observed data, our approach can yield treatment rules that retain higher average outcome, lower average loss, and higher accuracy. 

The remainder of the paper is organized as follows. In the next section, we formulate the general problem of decision making. We introduce our approach to generate accurate and interpretable treatment rules. In section \ref{sec: simulation study}, we conduct a simulation study and demonstrate the strength of our approach. In section \ref{sec: application}, we apply our approach to derive treatment rules for adjuvant chemotherapy in stage \rom{2} and \rom{3} colon cancer patients. We compare the estimated optimal treatment rules, the interpretable treatment rules and patients' self-selection treatment rules.

\section{Methodology} 
\label{sec: methodology}
In this section, we introduce our approach to estimate accurate and interpretable treatment rules. The approach requires three key steps, 1) we construct a conditional mean function of the outcome using a flexible model based on which the estimates of potential outcomes are derived; 2) given these estimates and a loss function, we estimate the optimal treatment for each individual; 3) we fit a model from a simple model family that is deemed interpretable to the estimate of the optimal treatment as a function of covariates, based on which the interpretable treatment rules are derived. We provide an overview of these three steps in \ref{algorithm: overview of approach to deriving the explainable treatment rule}. In the next section, we introduce the first step, relating decision-making to causal inference. 

\begin{algorithm}
    \caption{Overview of the approach to derive the interpretable treatment rules}
    \label{algorithm: overview of approach to deriving the explainable treatment rule}
    \begin{algorithmic}
        \State \textbf{1. Estimate potential outcomes} 
        \State \ \ a. Dataset $\mathcal{D}:\{(\boldsymbol{X}_{i},T_{i},Y_{i}):i=1,...,n\}$
        \State \ \ b. Choose a flexible model family to model $\mathbb{E}[Y\mid \boldsymbol{X},T]=f(\boldsymbol{X},T)$
        \State \ \ c. Select a $\widehat{f} = \text{argmin}_{\widehat{f} \in \mathcal{F}}L(\widehat{\tau}^{rct}, \sum_{i}^{n}\widehat{w}_{i}(\widehat{f}(\boldsymbol{X}_{i},1)-\widehat{f}(\boldsymbol{X}_{i},0)))$ \cite{shen2020estimating,shen2023RCTrep}, where $\widehat{w}_{i}$ is an estimate of weight for         
        \State \ \ \ \ \ \  individual $i$ to adjust for the covariate mismatch between two datasets, $L$ is a distance measurement. If data from 
        \State \ \ \ \ \ \  a RCT is not available, we can select $\widehat{f}$ using validation methods in \cite{schuler2017synth,shimoni2018benchmarking,JustCause,qian2011performance}.        
        \State \ \ d. Estimate joint potential outcomes for each individual $i$: $\widehat{\boldsymbol{\theta}}_{i}=(\widehat{\theta}_{i00},\widehat{\theta}_{i01},\widehat{\theta}_{i10},\widehat{\theta}_{i11})$
        \State \textbf{2. Estimate the optimal treatment rules} 
        \State \ \ a. Choose parameters for loss function $\boldsymbol{l}^{1}_{Y(1),Y(0)}=(l_{00}^{1},l_{01}^{1},l_{10}^{1},l_{11}^{1})$, and $\boldsymbol{l}^{0}_{Y(1),Y(0)}=(l_{00}^{0},l_{01}^{0},l_{10}^{0},l_{11}^{0})$
        \State \ \ b. Compute $\mathbb{L}(\widehat{f},1,\boldsymbol{X}_{i})=\widehat{\boldsymbol{\theta}}_{i}*\boldsymbol{l}^{1}_{Y(1),Y(0)}, \mathbb{L}(\widehat{f},0,\boldsymbol{X}_{i})=\widehat{\boldsymbol{\theta}}_{i}*\boldsymbol{l}^{0}_{Y(1),Y(0)}$ 
        \State \ \ c. Estimate the optimal treatment rules $\widehat{d^{*}}(\boldsymbol{X}_{i}) = \argmin_{t} \mathbb{L} (\widehat{f}, t, \boldsymbol{X}_{i}) $
        \State \textbf{3. Simplify the optimal treatment rules}
        \State \ \ a. Specify a simple model family to model $\mathbb{E}[\widehat{d^{*}} \mid \boldsymbol{X}]=r(\boldsymbol{X})$
        \State \ \ b. Specify a similarity measurement $D$
        \State \ \ c. Estimate $\widehat{r}=\argmin_{\boldsymbol{w} \in \mathcal{W}}\frac{1}{n} \sum_{i=1}^{n}  D\left(\widehat{d^{*}}(\boldsymbol{X}_{i}), \widehat{r}(\boldsymbol{X}_{i}) \right)$, 
        \State  \ \ d. Obtain the interpretable treatment rules $\widehat{d^{*}}_{s}(\boldsymbol{X}_{i})=\mathds{1}\{\widehat{r}(\boldsymbol{X}_{i}) > 0.5\}$
    \end{algorithmic}
\end{algorithm}

\subsection{Estimating potential outcomes}
Let $\boldsymbol{X}_{i}=(X_{i1},...,X_{ip}) \in \mathbb{R}^{p}$ denote the associated $p$-dimensional pretreatment covariates for individual $i$, $T_{i} \in \mathcal{T} = \{0,1\}$ denote the binary treatment indicator where 1 and 0 indicate the treatment and control, $Y_{i} \in  \{0,1\}$ denote the observed outcome, $Y_{i}(1)$ and $Y_{i}(0)$ denote the potential outcomes had the individual $i$ receive the treatment and control. We assume higher value of the outcome is desirable. Under the assumptions of unconfoundedness, overlap, and consistency \cite{imbens2015causal}, the expected value of $Y(t)$ for individuals $i$ with characteristics $\boldsymbol{X}_{i}=\boldsymbol{x}$ in a dataset can be estimated as the conditional mean of the outcome given $\boldsymbol{x}$ and $t$. Let $f: \mathcal{X}\times\mathcal{T}\mapsto (0,1)$ denote a model for conditional mean function of the outcome, so that 
$$\mathbb{E}[Y\mid\boldsymbol{X},T] \approx \widehat{f}(\boldsymbol{X},T).$$
For individual $i$ with $\boldsymbol{X}_{i}=\boldsymbol{x}$, potential outcomes are imputed using estimates of conditional mean $\widehat{f}(\boldsymbol{x},1)$ and $\widehat{f}(\boldsymbol{x},0)$. The most appropriate model $\widehat{f}$ from a set of candidates can be selected using our proposed method by comparing estimates of conditional average treatment effects to existing RCT results \cite{shen2023RCTrep}. If there is no existing RCT data, we can select the most appropriate model based on methods in \cite{schuler2017synth,shimoni2018benchmarking,JustCause} using synthetic truth of potential outcomes or select a model with the best outcome prediction accuracy \cite{qian2011performance}.
\subsection{Deriving optimal treatment rules}
\label{subsec: estimate optimal decision}
Consider a decision framework to derive the optimal treatment rules \cite{robert2007bayesian}. Let $d: \mathcal{X} \mapsto \mathcal{T}$ denote   a treatment rule, $L: (\mathcal{Y}(1), \mathcal{Y}(0)) \times \mathcal{T} \mapsto [0, +\infty)$ denote a loss function, $\mathbb{L} (\widehat{f}, t, \boldsymbol{X}_{i}) = \sum_{j,k\in\{0,1\}}l_{jk}^{t}\widehat{\theta}_{ijk}$ denote the estimate of expected loss of a treatment $t$ for $\boldsymbol{X}_{i}$, where $L(Y(1)=j, Y(0)=k, T=t)=l_{jk}^{t}$, $\widehat{\theta}_{ijk}$ is the estimate of joint probability of two potential outcomes for individual $\boldsymbol{X}_{i}$. See Appendix B for the computation of $\widehat{\theta}_{ijk}$. Then the optimal treatment $d^{*}(\boldsymbol{X}_{i})$ can be estimated by 
\begin{equation}
\label{equation: optimal treatment rule}
\widehat{d^{*}}(\boldsymbol{X}_{i}) = \argmin_{t} \mathbb{L} (\widehat{f}, t, \boldsymbol{X}_{i}) 
\end{equation}
In case $\widehat{f}$ is estimated using a Bayesian method, the probability that the estimated loss under the treatment lower than the estimated loss under the control can be easily quantified as $p(\widehat{d^{*}}(\boldsymbol{X}_{i})=1)=\frac{1}{D}\sum_{d=1}^{D}\mathds{1}\{\mathbb{L}(\widehat{f}^{(d)},1, \boldsymbol{X}_{i})<\mathbb{L}(\widehat{f}^{(d)},0, \boldsymbol{X}_{i}\}, d=1,...,D$ is the number of Monte Carlo samples from the posterior of $\widehat{f}$ \cite{klausch2018estimating,logan2019decision}.

\subsection{Simplifying treatment rules}
\label{subsec: step 2 summaries for the optimal decision rule within model constraints}
Since the optimal treatment rules are derived based on $\boldsymbol{X}_{i}$ that is often of high dimension, yet cost and interpretability consideration implies that only a few covariates should be used to construct treatment rules. Therefore, we demonstrate a procedure to simplify the estimated optimal treatment rules without considerably increasing average loss. Let $r: \mathcal{X} \mapsto (0,1)$ denote a simple model with parameters $\boldsymbol{w} \in \mathcal{W}$ that produces results close to $\widehat{d^{*}}(\boldsymbol{X}_{i})$ in a similarity measurement $D$. The specification of the model $r$ is deemed as our definition of interpretability. Under the problem formulation, we aim to model $\widehat{d^{*}}(\boldsymbol{X}_{i})$ using $r$ whose success probability depends on $\widehat{r}(\boldsymbol{X}_{i})$. The simplified treatment rules $\widehat{d^{*}}_{s}(\boldsymbol{X}_{i})$ are derived based on the best $\widehat{r}$ as follows:
\begin{equation}
\mathbb{E}[\widehat{d^{*}} \mid \boldsymbol{X}_{i}] \approx \widehat{r}(\boldsymbol{X}_{i}), \ \ \text{where} \ \ \widehat{r}  = \argmin_{\boldsymbol{w} \in \mathcal{W}} \frac{1}{n} \sum_{i=1}^{n}D(\widehat{d^{*}}(\boldsymbol{X}_{i}), \widehat{r}(\boldsymbol{X}_{i}))
\end{equation}
Then $\widehat{d^{*}}_{s}(\boldsymbol{X}_{i}) = \mathds{1}\{\widehat{r}(\boldsymbol{X}_{i}) > 0.5\}$. We choose cross-entropy as the similarity measurement. Minimizing the cross-entropy is equivalent to maximizing the log likelihood, thus $\widehat{r}$ is derived by maximizing the objective function \ref{equation:general objective function using cross-entropy loss} in Appendix C. If $\widehat{f}$ is estimated using a Bayesian method, $p(\widehat{d^{*}}(\boldsymbol{X}_{i})=1)$ is incorporated into the estimation. In our study, we use two model families as demonstration examples. The two model families are regression trees and logistic regressions. More details of deriving interpretable treatment rules using these two model families are provided in Appendix D. 

\section{Simulation study}
\label{sec: simulation study}
The simulation study aims to evaluate three optimal treatment rules estimated using different modelling techniques and procedures. These three treatment rules are $\widehat{d^{*}}(\boldsymbol{X}_{i})$, $\widehat{d^{*}}_{s'}(\boldsymbol{X}_{i})$, and $\widehat{d^{*}}_{s}(\boldsymbol{X}_{i})$. The set up for estimating these three optimal treatment rules are summarized in Table \ref{tab: set up of simulation}. We are interested in comparative performance of $\widehat{d^{*}}_{s}(\boldsymbol{X}_{i})$ and $\widehat{d^{*}}_{s'}(\boldsymbol{X}_{i})$. We aim to show that using the same model family, the estimated optimal treatment rules $\widehat{d^{*}}_{s}(\boldsymbol{X}_{i})$ derived from our proposed method is more accurate than $\widehat{d^{*}}_{s'}(\boldsymbol{X}_{i})$. 

\subsection{Data generation}
We reproduce the simulation settings from the relevant paper \cite{logan2019decision}, where we have one binary outcome $Y$, one binary treatment indicator $T$, and one pretreatment covariate vector $\boldsymbol{X}=(X_{A},...,X_{E},X_{a},...,X_{e},X_{Ca},X_{Cb})$. $X_{A}:X_{E}$ are five binary covariates with 0.5 success probability, $X_{a}:X_{e}$ are five ordinal covariates with four equally probable categories each, and $X_{Ca}$ and $X_{Cb}$ are two normally distributed random variables. Let $\mathbb{E}[Y \mid \boldsymbol{X},T]=f(\boldsymbol{X},T)$ denote the probability of the outcome taking value 1 conditional on $\boldsymbol{X}$ and $T$. The logit of $f$ is simulated according to the following eight scenarios, 
\begin{enumerate} 
    \item [A)] $0.5X_{C1} + 2(X_{B1}+X_{a3}*X_{A1})*T$
    \item [B)] $0.5X_{C1} + 2\{X_{B1}+X_{a3}*(X_{b2}+x_{b3})\}*T$
    \item [C)] $0.05(-X_{A1}+X_{B1})+\{(X_{a2}+X_{a3})+(X_{b2}+X_{b3})*X_{Ca}\}*T$
    \item [D)] $log log\{(X_{b3}+X_{c3})+5(X_{a2}+X_{a3}+X_{A1}X_{B1})*T+20\}^{2}$
    \item [E)] $(X_{A1}+X_{B1})+2*T$
    \item [F)] $0.5X_{A1}+0.5X_{B1}+2\mathds{1}\{X_{Ca}<5,X_{a}<2\}*T$
    \item [G)] $0.5X_{A1}+0.5X_{B1}+2\mathds{1}\{X_{Ca}<5,X_{Cb}<2\}*T$
    \item [H)] $0.5X_{Ca}+0.5X_{Cb}+2\mathds{1}\{X_{Ca}<-2,X_{Cb}>2\}*T$
\end{enumerate}
We simulate the probability of receiving the treatment as $P(T=1|\boldsymbol{X}) = logit^{-1}\left\{{\frac{\lambda(\tau(\boldsymbol{X})-\bar{\tau}(\boldsymbol{X}))}{\text{sd}(\tau(\boldsymbol{X}))}}\right\}$, where $\tau(\boldsymbol{X}) = \mathbb{E}[Y \mid \boldsymbol{X},T=1] - \mathbb{E}[Y \mid \boldsymbol{X},T=0]$ is conditional average treatment effects given $\boldsymbol{X}$, $\bar{\tau}(\boldsymbol{X})$ and $\text{sd}(\tau(\boldsymbol{X}))$ are sample mean and standard deviation of $\tau(\boldsymbol{X})$. We choose $\lambda=\log(3)$ to represent the magnitude of confounding, which leads to strong selectivity at individuals with large magnitude of treatment effects. 

\begin{table}[H]
\begin{center}
\caption{Set up for estimating three optimal treatment rules with a focus on the data for estimating treatment rules, models used for the construction of treatment rules, conditional mean function for these models, and rules for deriving the optimal treatment. }
\label{tab: set up of simulation}
\resizebox{\columnwidth}{!}{
\begin{tabular}{l l l l}
\hline
& $\widehat{d^{*}}(\boldsymbol{X}_{i})$ & $\widehat{d^{*}}_{s'}(\boldsymbol{X}_{i})$ & $\widehat{d^{*}}_{s}(\boldsymbol{X}_{i})$ \\ 
\hline
data & $(\boldsymbol{X}_{i},T_{i},Y_{i})$  & $(\boldsymbol{X}_{i},T_{i},Y_{i})$ & $(\boldsymbol{X}_{i},\widehat{d^{*}}(\boldsymbol{X}_{i}))$\\
funtion & $\mathbb{E}[Y \mid \boldsymbol{X},T] = f(\boldsymbol{X},T)$ & $\mathbb{E}[Y \mid \boldsymbol{X},T] = f(\boldsymbol{X},T)$ & $\mathbb{E}[\widehat{d^{*}} \mid \boldsymbol{X}] = r(\boldsymbol{X})$ \\
model & default-BART \cite{chipman2010bart} & $\bullet$ regression tree, $d_{max}=2$, $n_{obs}=5$ & $\bullet$  regression tree, $d_{max}=2, n_{obs}=5$\\
                     & $f(\boldsymbol{X},T)=\Phi(G(\boldsymbol{X},T))$           & \ \ \  $f(\boldsymbol{X},T)=\sum_{m=1}^{M} w_{m} \mathds{1}\{(\boldsymbol{X},T) \in R_{m}\}$ &  \ \ \ $r(\boldsymbol{X})=\sum_{m=1}^{M} w_{m}\mathds{1}\{\boldsymbol{X} \in R_{m}\}$ \\
                     & $G(\boldsymbol{X},T)=\sum_{j=1}^{m}g(\boldsymbol{X},T;T_{j},M_{j})$           & $\bullet$  logistic regression, & $\bullet$ logistic regression, \\
                     &            & \ \ \ $\text{logit}(f(\boldsymbol{X},T)) = \boldsymbol{w}\phi(\boldsymbol{X},T)$ & \ \ \ $\text{logit}(r(\boldsymbol{X})) = \boldsymbol{w}\phi(\boldsymbol{X})$ \\
                     &            & \ \ \ $\phi(\boldsymbol{X},T)=(X_{1},...,X_{p},T,X_{1}*T,..., X_{p}*T)$ &  \ \ \  $\phi(\boldsymbol{X})=(X_{1},...,X_{p})$ \\
rules & $\widehat{d^{*}}(\boldsymbol{X}_{i}) = \argmin_{t}\mathbb{L}(\widehat{f}, t, \boldsymbol{X}_{i})$ & $\widehat{d^{*}}_{s'}(\boldsymbol{X}_{i}) = \argmin_{t}\mathbb{L}(\widehat{f}, t, \boldsymbol{X}_{i})$ & $\widehat{d^{*}}_{s}(\boldsymbol{X}_{i}) = \mathds{1}\{\widehat{r}(\boldsymbol{X}_{i})>0.5\}$ \\
\hline
\end{tabular}}
\end{center}
\end{table}

\subsection{Simulation set-up} \label{subsec: simulation setup}
In this section, we introduce procedures for estimating three optimal treatment rules, i.e., $\widehat{d^{*}}(\boldsymbol{X}_{i})$, $\widehat{d^{*}}_{s'}(\boldsymbol{X}_{i})$, and $\widehat{d^{*}}_{s}(\boldsymbol{X}_{i})$, and evaluation metrics to compare these treatment rules. First, we define models for conditional mean function of the outcome, denoted as $\mathbb{E}[Y\mid \boldsymbol{X}, T]=f(\boldsymbol{X},T)$, to estimate potential outcomes. We use Bayesian additive regression trees (BART) with default settings of priors (i.e., $\alpha=0.95$, $\beta=2$, the number of tree $m=200$) since they are adequate and computationally efficient in most of cases \cite{chipman2010bart}. Parametes are estimated by Bayesian backfitting MCMC. We use a decision tree with maximum depth $d_{max}=2$ and minimun number of obervations in each node $n_{obs}=5$. Parameters of the decision tree are estimated by maximizing the objective function \ref{equation: objective functions for tree F2} in Appendix C. Estimates of potential outcomes using these two models are derived using imputed conditional mean under treatment and control. 

Next, based on estimates of potential outcomes derived from fitted BART model and the decision tree and a loss function, we derive two treatment rules $\widehat{d^{*}}(\boldsymbol{X}_{i})$ and $\widehat{d^{*}}_{s'}(\boldsymbol{X}_{i})$. For simplicity, we use the loss function in Table \ref{tab: additive loss function in our study} in Appendix E. Using this loss function, the optimal treatment rules are reformulated as an inequality based on estimates of conditional average treatment effects and a decision threshold. The threshold represents the weight of the loss due to the treatment over the loss due to the undesirable outcome. As the threshold increases, the optimal treatment rules tend to predict control as the optimal treatment. We set the decision threhold by $0\%,5\%, 10\%,15\%,20\%,..., 100\%$, and derive $\widehat{d^{*}}(\boldsymbol{X}_{i})$ and $\widehat{d^{*}}_{s'}(\boldsymbol{X}_{i})$ by the decision threhold. Lastly, we define a simple model family to estimate the conditional mean function of $\widehat{d^{*}}(\boldsymbol{X}_{i})$, denoted as $\mathbb{E}[\widehat{d^{*}} \mid \boldsymbol{X}]=r(\boldsymbol{X})$. We use a decision tree with $d_{max}=2, n_{obs}=5$ to model $r(\boldsymbol{X})$; $\widehat{r}(\boldsymbol{X})$ is derived by maximing the objective functions \ref{equation: objective functions for tree F2'}. $\widehat{d^{*}}_{s}(\boldsymbol{X}_{i})$ is derived by comparing $\widehat{r}(\boldsymbol{X}_{i})$ with 0.5. 

Similarly, we use logistic regressions to model the function $f(\boldsymbol{X},T)$ and the function $r(\boldsymbol{X})$, based on which $\widehat{d^{*}}_{s'}(\boldsymbol{X}_{i})$ and $\widehat{d^{*}}_{s}(\boldsymbol{X}_{i})$ are derived. We specify main effects of $\boldsymbol{X}$ and $T$ and interaction effects between $\boldsymbol{X}$ and $T$ in the conditional mean function $f(\boldsymbol{X},T)$. Parameters are estimated by maximizing the objective function \ref{equation: objective function for logistic regression F2} in Appendix C. We specify main effects of $\boldsymbol{X}$ in the conditional mean function $r(\boldsymbol{X})$. Parameters are estimated by maximizing the objective function in \ref{equation: objective function for logistic regression F2'} in Appendix D. Stochastic gradient descent is used to learn parameters of $f(\boldsymbol{X},T)$ and $r(\boldsymbol{X})$ with learning rate $\eta=0.01$ and the number of iterations $t=1000$. 

We compare performance of three estimated optimal treatment rules $d \in \{\widehat{d^{*}}(\boldsymbol{X}_{i})$, $\widehat{d^{*}}_{s'}(\boldsymbol{X}_{i}), \widehat{d^{*}}_{s}(\boldsymbol{X}_{i})\}$, using the following evaluation metrics, where $f$ is true DGM, 1) the average loss $R(d) = \frac{1}{n}\sum_{i} \mathbb{L}(f, d(\boldsymbol{X}_{i}),\boldsymbol{X}_{i})$ \cite{robert2007bayesian}; 2) the average outcome $V(d) = \frac{1}{n}\sum_{i} f(\boldsymbol{X}_{i},d(\boldsymbol{X}_{i}))$ \cite{qian2011performance}; 3) accuracy, precision, and recall of treatment rules, where precision is the proportion of prediction of the optimal treatment being 1 that is actually correct, defined as $\sum\mathds{1}\{d(\boldsymbol{X}_{i})=1,d^{*}(\boldsymbol{X}_{i})=1\}/ \sum\mathds{1}\{d(\boldsymbol{X}_{i})=1\}$, and recall is the proportion of individuals whose optimal treatment is 1 that are correctly identified, defined as $\sum\mathds{1}\{d(\boldsymbol{X}_{i})=1,d^{*}(\boldsymbol{X}_{i})=1\}/ \sum\mathds{1}\{d^{*}(\boldsymbol{X}_{i})=1\}$. We generate a population of size 10000, from which we draw 100 samples each of size $n=1000$. Average $R$, $V$, accuracy, precision, and recall and their standard errors across 100 samples are obtained. 

\subsection{Results} 
The results of $R$ and $V$ are present in Figure \ref{fig: simulation_loss_outcome_LReg} and Figure \ref{fig: simulation_loss_outcome_TREE_depth1} in Appendix G, showing that $\widehat{d^{*}}(\boldsymbol{X}_{i})$ has the lowest $R$ and the highest $V$ as expected. Compared to $\widehat{d^{*}}_{s'}(\boldsymbol{X}_{i})$, $\widehat{d^{*}}_{s}(\boldsymbol{X}_{i})$ has lower $R$ and higher $V$; $\widehat{d^{*}}_{s}(\boldsymbol{X}_{i})$ has smaller variability of $R$ and $V$ than those of $\widehat{d^{*}}_{s'}(\boldsymbol{X}_{i})$. As the decision threhold increases, all treatment rules tend to have comparable $R$ and $V$. The accuracy of treatment rules is shown in Figure \ref{fig: simulation_BART_TREE1_acc_precision_recall} and Figure \ref{fig: simulation_BART_LReg_acc_precision_recall} in Appendix G. The results indicate that overall $\widehat{d^{*}}(\boldsymbol{X}_{i})$ is the most accurate treatment rule; $\widehat{d^{*}}_{s}(\boldsymbol{X}_{i})$ is comparable to $\widehat{d^{*}}(\boldsymbol{X}_{i})$ and is more accurate than $\widehat{d^{*}}_{s'}(\boldsymbol{X}_{i})$. To further investigate the type of errors that three treatment rules make, we examine precision and recall. The results in Figure \ref{fig: simulation_BART_TREE1_acc_precision_recall} and Figure \ref{fig: simulation_BART_LReg_acc_precision_recall} in Appendix G show that overall $\widehat{d^{*}}(\boldsymbol{X}_{i})$ and $\widehat{d^{*}}_{s}(\boldsymbol{X}_{i})$ have slighly higher precision and considerably higher recall than $\widehat{d^{*}}_{s'}(\boldsymbol{X}_{i})$, implying that $\widehat{d^{*}}(\boldsymbol{X}_{i})$ and $\widehat{d^{*}}_{s}(\boldsymbol{X}_{i})$ are more powerful to identify individuals who can benefit from the treatment. The relatively low recall of $\widehat{d^{*}}_{s'}(\boldsymbol{X}_{i})$ can lead to under-treatment. The higher recall of $\widehat{d^{*}}_{s}(\boldsymbol{X}_{i})$ may due to the variance reduction in fitted values used for estimating parameters in $\widehat{d^{*}}_{s}(\boldsymbol{X}_{i})$ compared to observed data used for estimating parameters in $\widehat{d^{*}}_{s'}(\boldsymbol{X}_{i})$, where $\widehat{d^{*}}_{s}(\boldsymbol{X}_{i})$ is a summary of the imputed values from $\widehat{d^{*}}(\boldsymbol{X}_{i})$ while $\widehat{d^{*}}_{s'}(\boldsymbol{X}_{i})$ is a summary of noisy observations \cite{woody2020model}. Similar results can be found for $f$ and $r$ which are estimated using binary trees with $d_{max} = 3,...,10$, and can be reproduced using provided codes. Thus we do not present results in this paper. 

\begin{figure}[h]
    \centering
    \includegraphics[width=1\textwidth]{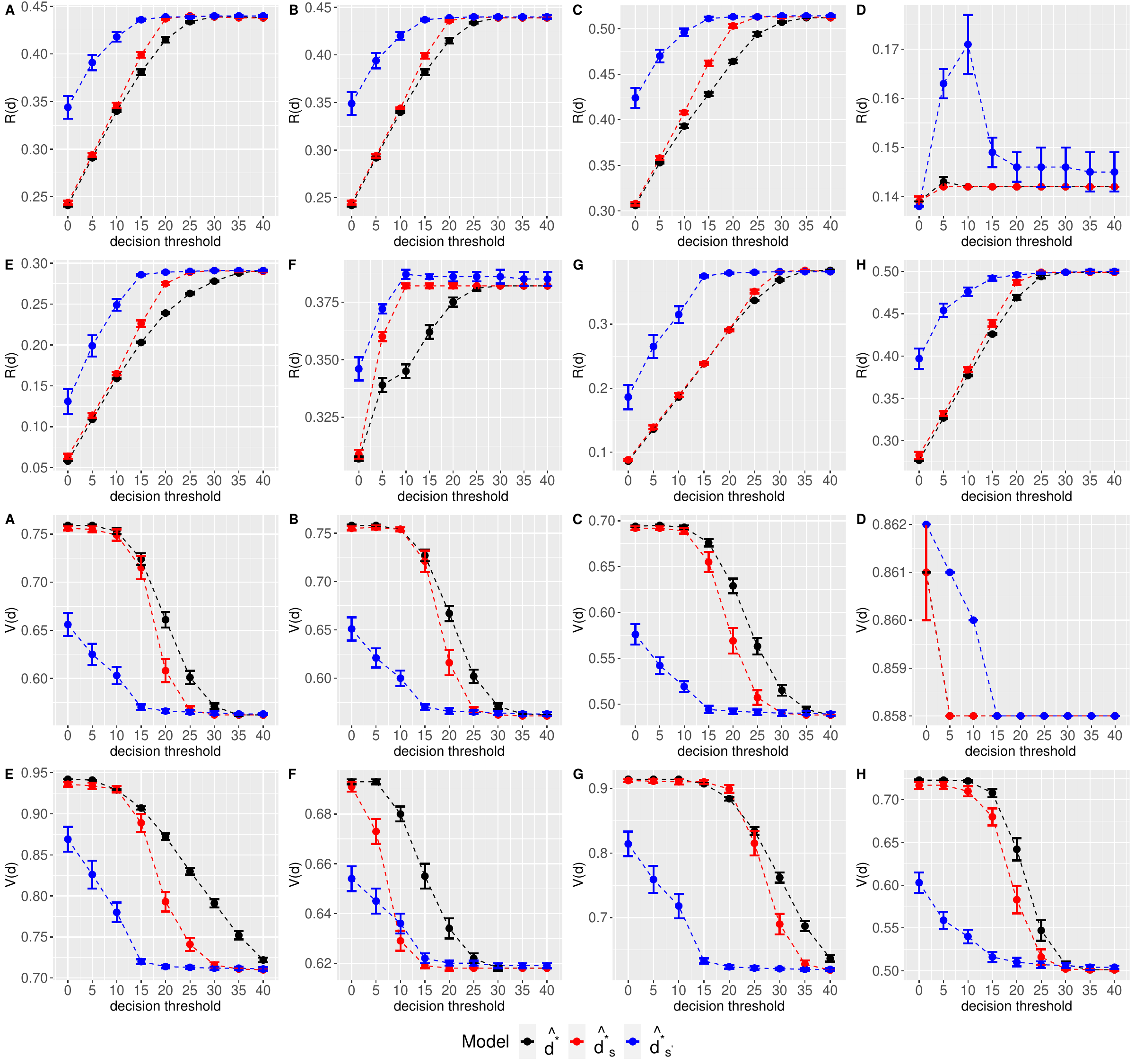}
    \caption{Average loss $R$ and average outcome $V$ of three treatment rules $\widehat{d^{*}}(\boldsymbol{X}_{i})$, $\widehat{d^{*}}_{s'}(\boldsymbol{X}_{i})$, and $\widehat{d^{*}}_{s}(\boldsymbol{X}_{i})$, where $f(\boldsymbol{X},T)$ for estimating  $\widehat{d^{*}}_{s'}(\boldsymbol{X}_{i})$ and $r(\boldsymbol{X})$ for estimating $\widehat{d^{*}}_{s}(\boldsymbol{X}_{i})$ are constructed using logistic regressions, the decision threhold denotes the weight of the loss due to the treatment over the loss due to the undesirable outcome.}
    \label{fig: simulation_loss_outcome_LReg} 
\end{figure}

\clearpage

\section{Application to real-world data}
\label{sec: application}
In this section, we apply our approach to derive both accurate and interpretable treatment rules for adjuvant chemotherapy (ACT) in stage II and III colon cancer (CC) as an illustrative example. We compare among these treatment rules and against patients' self-selection treatment rules in terms of average loss and average outcome. 

\subsection{Colon cancer registry data}\
The Netherlands Cancer Registry is a registry containing all cancer types in the Netherlands. A total of $n=27,057$ stage II and III CC patients who underwent curative surgery between 2006 and 2016 from the registry were selected to estimate potential outcomes under surgery only and surgery with ACT.  The outcome is 5-year OS, defined as time from 30 days after the start date of the surgery to death from any cause or censoring for patients who were still alive at the time of follow-up cutoff. Indication of ACT administration is defined as presence of any registration in the data of ACT after the start date of the surgery. The detailed description of data is provided in Table \ref{tab: summary statistics of the study population} in Appendix F. 

\subsection{Method}  \label{subsec: app method}
We perform the following steps, 1) we use the default probit BART for the binary outcomes to model the conditional mean function of the outcome $f$. The model regresses the outcome on 15 covariates in Table \ref{tab: summary statistics of the study population} , a binary treatment indicator of ACT, and the propensity score for receiving ACT given these covariates. The study used the default settings to train the outcome model, namely, $\alpha=2$, $\beta=0.95$, $m=200$ for the number of trees. The number of MCMC iterations was 1100, in which 100 iterations were treated as burn-in steps and were dropped and 1000 iterations were returned to estimate the posterior. The same settings are used to model the propensity score. 2) we employ the loss function in Table \ref{tab: additive loss function in our study}, based on which the estimated optimal treatment rules $\widehat{d^{*}}(\boldsymbol{X}_{i})$ are derived. The loss function leads to a decision threshold representing the weight of the loss due to ACT over the loss due to death. As the decision threhold increases, the optimal treatment  rules tend to predict no ACT as the optimal treatment. Four settings of the decision threhold are specified, namely, 0\%, 5\%,10\%, and 15\%. 3) we define an interpretable model family as a binary tree with maximum depth 2 and minimum number of observations in nodes 50 to model $r(\boldsymbol{X})$. The simple treatment rules $\widehat{d^{*}}_{s}(\boldsymbol{X}_{i})$ are derived. Additionally, the observed treatment rules are derived from the observed treatment indicator, denoted by $\widehat{d^{*}}_{obs}(\boldsymbol{X}_{i})=T_{i}$. Three treatment rules are compared in terms of average loss $R$ and average outcome $V$. We consider $\widehat{f}$ estimated using BART as an accurate estimate of $f$ and we use the MC sample mean of $\mathbb{L}(\widehat{f}^{(d)}, d(\boldsymbol{X}_{i}), \boldsymbol{X}_{i})$ for computation of $R$ and MC sample mean of $\widehat{f}^{(d)}(\boldsymbol{X}_{i}, d(\boldsymbol{X}_{i}))$ for computation of $V$, where $d=1,...,1000$. 

\subsection{Results}
\begin{figure}[ht]
    \centering
    \includegraphics[scale=0.3]{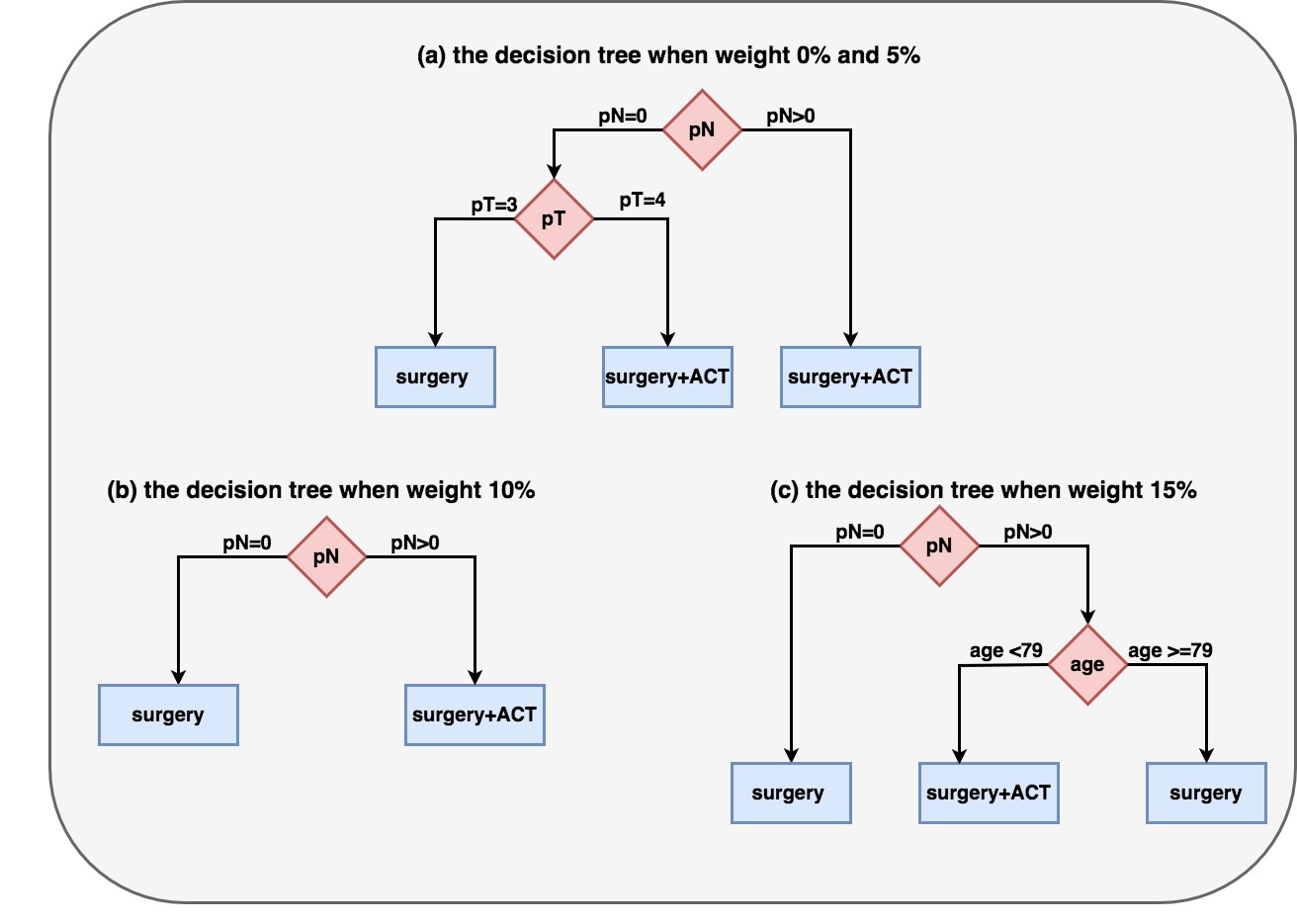}
    \caption{The decision trees for ACT in stage II and III CC patients when the decision threshold representing the weight of the loss due to  ACT over the loss due to death is 0\%, 5\%, 10\%, and 15\% in tree (a), (b), and (c), respectively.}
    \label{fig: decisiont trees}
\end{figure}

Three decision trees under four settings of the decision threhold are generated and are present in Figure \ref{fig: decisiont trees}, where the decision trees under the decision threhold 0\% and 5\% are the same. The results indicate that 1) in general, pN is the most important covariate to predict the optimal treatment; 2) when the decision threhold is less than 10\%, in addition to pN, pT is the most important covariate to predict the optimal treatment within the subset of patients with $\text{pN}=0$; 3) when the decision threhold is 15\%, in addition to pN, age is the most important covariate to predict the optimal treatment within the subset of patients with $\text{pN}>0$. By analyzing the decision trees, we can identify the most important covariate for making the optimal treatment by different decision threholds. The estimates of $R$ and $V$ of $\widehat{d^{*}}(\boldsymbol{X}_{i})$, $\widehat{d^{*}}_{s}(\boldsymbol{X}_{i})$ and $\widehat{d^{*}}_{obs}(\boldsymbol{X}_{i})$ are present in Table \ref{tab: average loss of optimal and tree rules}. The results show that $R$ and $V$ of $\widehat{d^{*}}(\boldsymbol{X}_{i})$ and $\widehat{d^{*}}_{s}(\boldsymbol{X}_{i})$ are similar; $\widehat{d^{*}}_{obs}(\boldsymbol{X}_{i})$ has the lowest $V$ and highest $R$, pointing to the potential of using $\widehat{d^{*}}(\boldsymbol{X}_{i})$ and $\widehat{d^{*}}_{s}(\boldsymbol{X}_{i})$ to improve patients' self-selected treatment decisions. 

\begin{table} [h]
\centering
\caption{Average loss $R$ and average outcome $V$ for $\widehat{d^{*}}(\boldsymbol{X}_{i})$, $\widehat{d^{*}}_{s}(\boldsymbol{X}_{i})$, and $\widehat{d^{*}}_{obs}(\boldsymbol{X}_{i})$. Bold numbers indicate lower $R$ and higher $V$ between $\widehat{d^{*}}_{s}(\boldsymbol{X}_{i})$ and $\widehat{d^{*}}_{obs}(\boldsymbol{X}_{i})$. }
\label{tab: average loss of optimal and tree rules}
\begin{tabular} {l l l l l l l}
\hline
threshold & $R(\widehat{d^{*}})$ & $R(\widehat{d^{*}}_{s})$ & $R(\widehat{d^{*}}_{obs})$ & $V(\widehat{d^{*}})$ & $V(\widehat{d^{*}}_{s})$ & $V(\widehat{d^{*}}_{obs})$ \\ 
\hline
0\% & 0.305 &  \textbf{0.305} & 0.340 & 0.695 & \textbf{0.695} & 0.660 \\
5\%&0.334 & \textbf{0.334} & 0.357 & 0.694 & \textbf{0.695} & 0.660 \\
10\%& 0.360 & \textbf{0.362} & 0.375 & 0.688 & \textbf{0.689} & 0.660 \\
15\%&0.379 & \textbf{0.385} & 0.392 & 0.655 & \textbf{0.676} & 0.660 \\
\hline
\end{tabular}
\end{table}

\clearpage

\section{Conclusion and future work} 
\label{sec:conclusion and future work} 
In our study, we propose an approach to derive both accurate \textit{and} intepretable treatment rules. Firstly, we use a flexible model to estimate the optimal treatment rules for each individual. Then we specify a simple model as a definition of intepretability, from which we derive the interpretable treatment rules that are close to the optimal treatment rules. In the simulation study, we demonstrate that compared to deriving treatment rules from the same model fitted to observed data directly, our approach can reduce average loss, improve average outcome, and can be more powerful in identifying individuals who can benefit from the treatment. The method is illustrated by analyzing the CC registry data. The interpretable treatment rules of ACT are derived and are compared to the estimated optimal treatment rules and the patients' self-selection treatment rules, showing that these interpretable treatment rules can potentially improve patients' self-selection treatment decisions. 

To address the ambiguity of interpretating the general loss function in decision making at various levels, an additive form of the loss function in Appendix E, which consists of two loss values (i.e., $c_{t}$ and $c_{n}$) is introduced. These two loss values have practical implications for different decision makers. The optimal treatment rules are reformulated as an inequality based on these two loss values and estimates of conditional average treatment effects. When considering benefit and safety of the treatment, the optimal treatment rules can be reformulated as $\widehat{d^{*}}(\boldsymbol{X}_{i})=\mathds{1}\{\widehat{\tau}(\boldsymbol{X}_{i})/c_{t}>1/c_{n}\}$, where $c_{t}$ can be regarded as the probability of presence of a side effect, and $c_{n}$ can be regarded as a threshold to determine the relative importance of effectiveness over safety. A higher value of $c_{n}$ corresponds to a stronger emphasis on effectiveness. When considering value and cost of the treatment, the optimal treatment rules can be reformulated as $\widehat{d^{*}}(\boldsymbol{X}_{i})=\mathds{1}\{c_{t}/\widehat{\tau}(\boldsymbol{X}_{i})<c_{n}\}$, where $c_{t}$ can be regarded as the monetary cost of the treatment, $c_{t}/\widehat{\tau}(\boldsymbol{X}_{i})$ can be regarded as the incremental cost-effective ratio (ICER), and $c_{n}$ can be regarded as a threhold. The treatment rule implies that when the ICER for unit $\boldsymbol{X}_{i}$ is smaller than the threshold, the unit will receive the treatment.  As $c_{n}$ increases, $\widehat{d^{*}}$ tends to predict the treatment as the optimal treatment. In both cases, $c_{t}$ can be a function of $\boldsymbol{X}$ and can be modeled using available data. Future work could explore how to apply our approach using this loss function to real use-cases. 

In the future, more research could be conducted to examine which factors can explain the discrepancy between observed treatments and the predicted optimal treatment and whether and how these factors are correlated with outcomes. Additionally, further research could identify and rank subgroups with large variability of the optimal treatment estimated from the Bayesian method. Experiments can be designed to explore potential factors that may reduce the variability. Lastly, although our approach can be easily extended to the problem of multi-arm treatments, extending our approach to continuous and time-to-event outcomes is challenging, and further study could be done to investigate how to generalize our approach to such outcomes. 

\clearpage

\section*{Acknowledgments}
The authors thank the registration team of the Netherlands Comprehensive Cancer Organisation (IKNL) for the collection of data for the Netherlands Cancer Registry, the Prospective National Colon Rectum Cancer (PLCRC) working group for the collection of Patient Reported Outcome data as well as IKNL staff for scientific advice.

\subsection*{Author contributions}
SLJ carried out the statistical analyses. SLJ and MCK drafted the manuscript. SLJ participated in the interpretation of the statistical analyses. All authors critically read the drafts of this paper, and read and approved its final version.

\subsection*{Ethics approval and consent to participate}
Permission to access the Netherlands Cancer registry (NCR) and the databases of Statistics Netherlands (CBS) was granted by both IKNL (Netherlands Comprehensive Cancer Care Organisation) and the CBS (Statistics Netherlands). According to the Dutch Medical Research Involving Human Subjects Act, this study did not require medical-ethical approval, as was confirmed in writing by the medical ethical committee of the AMC, 10 July 2013. We followed the ethical principles for medical research involving human subjects as laid down in the Declaration of Helsinki and adopted by the World Medical Association (WMA Declaration of Helsinki, 2000).

\subsection*{Conflict of interest}
The authors declare no potential conflict of interests.

\subsection*{Availability of data and replication codes}
The data that support the findings of this study are available from IKNL and CBS but restrictions apply to the availability of these data, which were used under license for the current study, and so are not publicly available. The replication codes of the simulation study are available on https://github.com/duolajiang/Making-accurate-and-interpretable-decisions-for-binary-outcome. The R package IntpTrtRule is developed for deriving a simple decision tree using our approach, and is available on https://github.com/duolajiang/IntpTrtRule. 


\bibliographystyle{unsrt}  
\bibliography{manuscript}  

\begin{thebibliography}{10}

\bibitem{collins2015new}
Francis~S Collins and Harold Varmus.
\newblock A new initiative on precision medicine.
\newblock {\em New England journal of medicine}, 372(9):793--795, 2015.

\bibitem{qian2011performance}
Min Qian and Susan~A Murphy.
\newblock Performance guarantees for individualized treatment rules.
\newblock {\em Annals of statistics}, 39(2):1180, 2011.

\bibitem{imbens2015causal}
Guido~W Imbens and Donald~B Rubin.
\newblock {\em Causal inference in statistics, social, and biomedical
  sciences}.
\newblock Cambridge University Press, 2015.

\bibitem{lipkovich2017tutorial}
Ilya Lipkovich, Alex Dmitrienko, and Ralph B~D'Agostino~Sr.
\newblock Tutorial in biostatistics: data-driven subgroup identification and
  analysis in clinical trials.
\newblock {\em Statistics in medicine}, 36(1):136--196, 2017.

\bibitem{kosorok2019precision}
Michael~R Kosorok and Eric~B Laber.
\newblock Precision medicine.
\newblock {\em Annual review of statistics and its application}, 6:263--286,
  2019.

\bibitem{xu2020optimal}
Yuejia Xu, Angela~M Wood, Michael~J Sweeting, David~J Roberts, and Brian~DM
  Tom.
\newblock Optimal individualized decision rules from a multi-arm trial: A
  comparison of methods and an application to tailoring inter-donation
  intervals among blood donors in the uk.
\newblock {\em Statistical Methods in Medical Research}, page 0962280220920669,
  2020.

\bibitem{logan2019decision}
Brent~R Logan, Rodney Sparapani, Robert~E McCulloch, and Purushottam~W Laud.
\newblock Decision making and uncertainty quantification for individualized
  treatments using bayesian additive regression trees.
\newblock {\em Statistical methods in medical research}, 28(4):1079--1093,
  2019.

\bibitem{klausch2018estimating}
Thomas Klausch, Peter van~de Ven, Tim van~de Brug, Mark~A van~de Wiel, and
  Johannes Berkhof.
\newblock Estimating bayesian optimal treatment regimes for dichotomous
  outcomes using observational data.
\newblock {\em arXiv preprint arXiv:1809.06679}, 2018.

\bibitem{foster2011subgroup}
Jared~C Foster, Jeremy~MG Taylor, and Stephen~J Ruberg.
\newblock Subgroup identification from randomized clinical trial data.
\newblock {\em Statistics in medicine}, 30(24):2867--2880, 2011.

\bibitem{zhang2012estimating}
Baqun Zhang, Anastasios~A Tsiatis, Marie Davidian, Min Zhang, and Eric Laber.
\newblock Estimating optimal treatment regimes from a classification
  perspective.
\newblock {\em Stat}, 1(1):103--114, 2012.

\bibitem{zhang2012robust}
Baqun Zhang, Anastasios~A Tsiatis, Eric~B Laber, and Marie Davidian.
\newblock A robust method for estimating optimal treatment regimes.
\newblock {\em Biometrics}, 68(4):1010--1018, 2012.

\bibitem{zhao2012estimating}
Yingqi Zhao, Donglin Zeng, A~John Rush, and Michael~R Kosorok.
\newblock Estimating individualized treatment rules using outcome weighted
  learning.
\newblock {\em Journal of the American Statistical Association},
  107(499):1106--1118, 2012.

\bibitem{zhou2023offline}
Zhengyuan Zhou, Susan Athey, and Stefan Wager.
\newblock Offline multi-action policy learning: Generalization and
  optimization.
\newblock {\em Operations Research}, 71(1):148--183, 2023.

\bibitem{hitsch2018heterogeneous}
Guenter~J. Hitsch, Sanjog Misra, and Walter Zhang.
\newblock Heterogeneous treatment effects and optimal targeting policy
  evaluation.
\newblock {\em Available at SSRN 3111957}, February 28, 2023.

\bibitem{chatton2020g}
Arthur Chatton, Florent Le~Borgne, Cl{\'e}mence Leyrat, Florence Gillaizeau,
  Chlo{\'e} Rousseau, Laetitia Barbin, David Laplaud, Maxime L{\'e}ger, Bruno
  Giraudeau, and Yohann Foucher.
\newblock G-computation, propensity score-based methods, and targeted maximum
  likelihood estimator for causal inference with different covariates sets: a
  comparative simulation study.
\newblock {\em Scientific reports}, 10(1):1--13, 2020.

\bibitem{shen2020estimating}
Lingjie Shen, Erik Visser, Hans de~Wilt, Henk Verheul, Felice van Erning, Gijs
  Geleijnse, and Maurits Kaptein.
\newblock Estimating the effect of adjuvant chemo-therapy for colon-cancer
  using registry data: a method comparison and validation.
\newblock 2020.

\bibitem{shen2023RCTrep}
Gijs~Geleijnse Lingjie~Shen and Maurits Kaptein.
\newblock Rctrep: An r package for the validation of estimates of average
  treatment effects.

\bibitem{dorie2019automated}
Vincent Dorie, Jennifer Hill, Uri Shalit, Marc Scott, Dan Cervone, et~al.
\newblock Automated versus do-it-yourself methods for causal inference: Lessons
  learned from a data analysis competition.
\newblock {\em Statistical Science}, 34(1):43--68, 2019.

\bibitem{hastie2009elements}
Trevor Hastie, Robert Tibshirani, and Jerome Friedman.
\newblock {\em The elements of statistical learning: data mining, inference,
  and prediction}.
\newblock Springer Science \& Business Media, 2009.

\bibitem{woody2020model}
Spencer Woody, Carlos~M Carvalho, and Jared~S Murray.
\newblock Model interpretation through lower-dimensional posterior
  summarization.
\newblock {\em Journal of Computational and Graphical Statistics}, pages 1--9,
  2020.

\bibitem{peltola2018local}
Tomi Peltola.
\newblock Local interpretable model-agnostic explanations of bayesian
  predictive models via kullback-leibler projections.
\newblock {\em arXiv preprint arXiv:1810.02678}, 2018.

\bibitem{piironen2018projective}
Juho Piironen, Markus Paasiniemi, and Aki Vehtari.
\newblock Projective inference in high-dimensional problems: prediction and
  feature selection.
\newblock {\em arXiv preprint arXiv:1810.02406}, 2018.

\bibitem{namkoong2020distilled}
Hongseok Namkoong, Samuel Daulton, and Eytan Bakshy.
\newblock Distilled thompson sampling: Practical and efficient thompson
  sampling via imitation learning.
\newblock {\em arXiv preprint arXiv:2011.14266}, 2020.

\bibitem{lundberg2020local}
Scott~M Lundberg, Gabriel Erion, Hugh Chen, Alex DeGrave, Jordan~M Prutkin,
  Bala Nair, Ronit Katz, Jonathan Himmelfarb, Nisha Bansal, and Su-In Lee.
\newblock From local explanations to global understanding with explainable ai
  for trees.
\newblock {\em Nature machine intelligence}, 2(1):56--67, 2020.

\bibitem{moncada2021explainable}
Arturo Moncada-Torres, Marissa~C van Maaren, Mathijs~P Hendriks, Sabine
  Siesling, and Gijs Geleijnse.
\newblock Explainable machine learning can outperform cox regression
  predictions and provide insights in breast cancer survival.
\newblock {\em Scientific reports}, 11(1):6968, 2021.

\bibitem{schuler2017synth}
Alejandro Schuler, Ken Jung, Robert Tibshirani, Trevor Hastie, and Nigam Shah.
\newblock Synth-validation: Selecting the best causal inference method for a
  given dataset.
\newblock {\em arXiv preprint arXiv:1711.00083}, 2017.

\bibitem{shimoni2018benchmarking}
Yishai Shimoni, Chen Yanover, Ehud Karavani, and Yaara Goldschmnidt.
\newblock Benchmarking framework for performance-evaluation of causal inference
  analysis.
\newblock {\em arXiv preprint arXiv:1802.05046}, 2018.

\bibitem{JustCause}
{Maximilian Franz, Florian Wilhelm, Tanmay Kulkarni }.
\newblock Justcause: Comparing methods for causality analysis in a fair and
  just way.

\bibitem{robert2007bayesian}
Christian Robert.
\newblock {\em The Bayesian choice: from decision-theoretic foundations to
  computational implementation}.
\newblock Springer Science \& Business Media, 2007.

\bibitem{chipman2010bart}
Hugh~A Chipman, Edward~I George, Robert~E McCulloch, et~al.
\newblock Bart: Bayesian additive regression trees.
\newblock {\em The Annals of Applied Statistics}, 4(1):266--298, 2010.

\bibitem{breiman1984classification}
Leo Breiman, Jerome Friedman, Charles~J Stone, and Richard~A Olshen.
\newblock {\em Classification and regression trees}.
\newblock CRC press, 1984.

\bibitem{tsuruoka2009stochastic}
Yoshimasa Tsuruoka, Jun’ichi Tsujii, and Sophia Ananiadou.
\newblock Stochastic gradient descent training for l1-regularized log-linear
  models with cumulative penalty.
\newblock In {\em Proceedings of the Joint Conference of the 47th Annual
  Meeting of the ACL and the 4th International Joint Conference on Natural
  Language Processing of the AFNLP}, pages 477--485, 2009.

\end{thebibliography}

\clearpage

\appendix{Appendix A: Motivation}
\label{app: motivation}

\begin{figure}[h]
    \centering
    \includegraphics[scale=0.6]{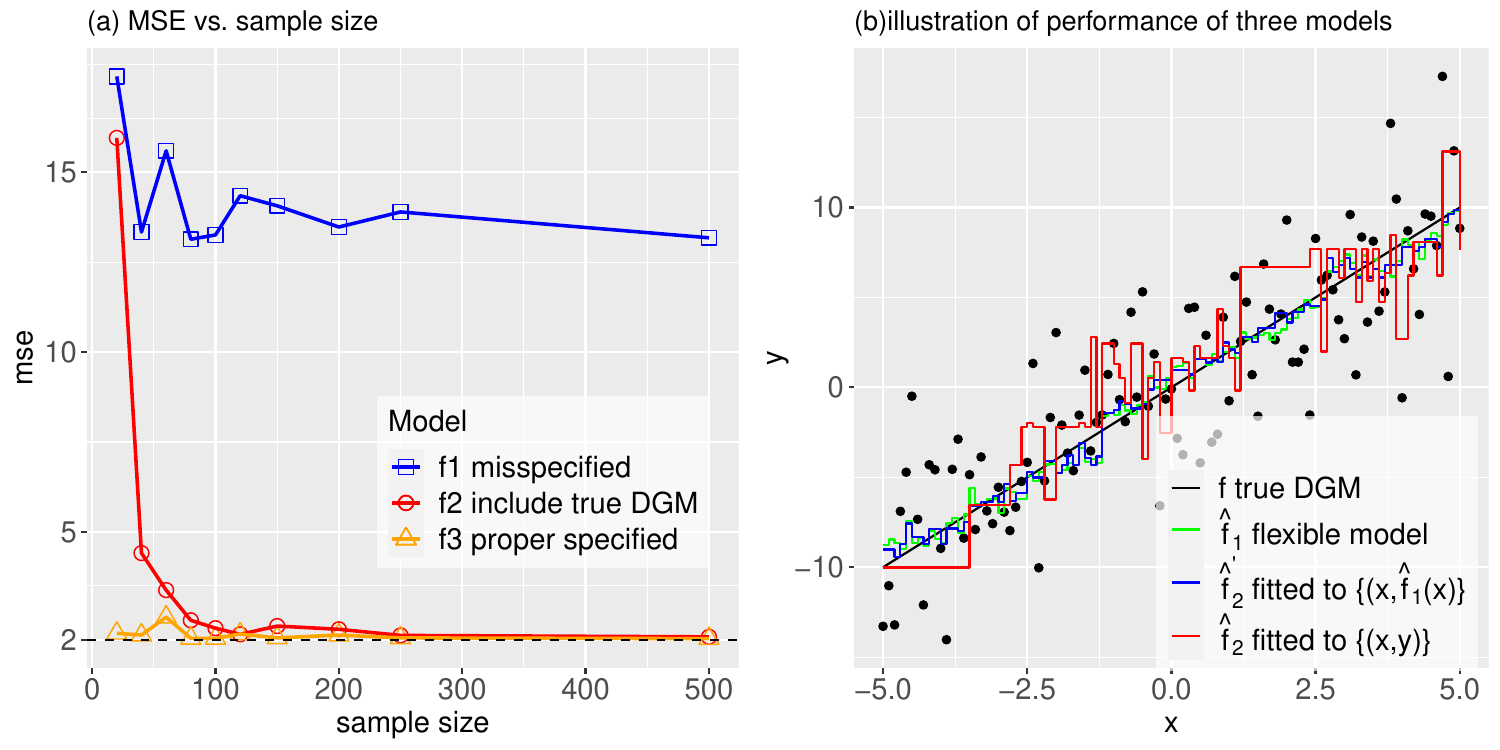}
    \caption{a) The effect of model space on bias. The variance of error term of the true data generation mechanism (DGM) is 2. The function form of $f_{1}$ and $f_{2}$ is correctly specified and the function form of $f_{3}$ is misspecified. $f_{2}$ includes the true function form. As the sample size increases, mean square error (MSE) of $\widehat{f}_{2}$ is approaximately equal to 2 while MSE of $\widehat{f}_{3}$ does not converge to 2, which may imply that the error of $\widehat{f}_{2}$ is from the variability of the true DGM while the error of $\widehat{f}_{3}$ is from the irreducible bias. b) Intuitive advantage of our proposed appoach.}
    \label{fig: motivation} 
\end{figure}
\clearpage

\appendix{Appendix B: Estimates of joint distribution of potential outcomes}
\label{Appendix B:joint distribution of potential outcomes}
\begin{equation}
\begin{split}
\hat{\theta}_{i11}&=\hat{p}(Y(1)=1,Y(0)=1\mid \boldsymbol{X}_{i}) = \rho\sqrt{\hat{\theta}_{i1.}(1-\hat{\theta}_{i1.})\hat{\theta}_{i.1}(1-\hat{\theta}_{i.1})}+\hat{\theta}_{i1.}\hat{\theta}_{i.1} \\
\hat{\theta}_{i10}&=\hat{p}(Y(1)=1,Y(0)=0\mid \boldsymbol{X}_{i}) =\hat{\theta}_{i1.}-\hat{\theta}_{i11}, \hat{\theta}_{i01} = \hat{\theta}_{i.1}- \hat{\theta}_{i11} \\
\hat{\theta}_{i00}&=\hat{p}(Y(1)=0,Y(0)=0\mid \boldsymbol{X}_{i}) = 1-\hat{\theta}_{i11}-\hat{\theta}_{i10}-\hat{\theta}_{i01}\\
\hat{\theta}_{i1.}&=\hat{p}(Y(1)=1\mid \boldsymbol{X}_{i}) =\widehat{f}(\boldsymbol{X}_{i},1) \\
\hat{\theta}_{i.1}&=\hat{p}(Y(0)=1\mid \boldsymbol{X}_{i}) =\widehat{f}(\boldsymbol{X}_{i},0)
\end{split}
\end{equation}
where $\rho$ is pairwise correlation between $Y(1)$ and $Y(0)$. When $\widehat{f} \sim p(\widehat{f}|X,T,Y)$, $\hat{\theta}_{ijk}, j,k \in \{0,1\}$ is estimated as the average of $\hat{\theta}_{ijk}^{(d)}$ derived from samples $\widehat{f}^{(1)},\widehat{f}^{(2)},..., \widehat{f}^{(d)}$. 

\clearpage

\appendix{Appendix C: Objective functions}
\label{Appendix C: objective function}

\begin{equation}
\label{equation:general objective function using cross-entropy loss}
\widehat{r} = \argmax_{\boldsymbol{w} \in \mathcal{W}}  \frac{1}{n}\sum_{i=1}^{n}p(\widehat{d^{*}}(\boldsymbol{X}_{i})=1)\log r(\boldsymbol{X}_{i};\boldsymbol{w})+(1-p(\widehat{d^{*}}(\boldsymbol{X}_{i})=1))\log (1-r(\boldsymbol{X}_{i};\boldsymbol{w}))
\end{equation}

\begin{equation}
\label{equation: objective functions for tree F2}
\widehat{f} = \argmax_{\mathcal{W},\mathcal{R}} \frac{1}{n}\sum_{m=1}^{M}\sum_{(\boldsymbol{x}_{i},t_{i}) \in R_{m}}y_{i}\log w_{m}+(1-y_{i})\log (1-w_{m}))
\end{equation}

\begin{equation}
\label{equation: objective function for logistic regression F2}
\widehat{f} = \argmax_{\mathcal{W}} \frac{1}{n}\sum_{i=1}^{n}y_{i}\log f(\boldsymbol{x}_{i},t_{i};\boldsymbol{w})+(1-y_{i})\log (1-f(\boldsymbol{x}_{i},t_{i};\boldsymbol{w})). 
\end{equation}

\clearpage

\appendix{Appendix D: Deriving interpretable treatment rules using two model families}
\label{Appendix D: deriving interpretable treatment rules using two model families}

\section{Deriving interpretable treatment rules using a decision tree}
\label{subsec: simplify using binary regression trees}
Since the tree structure is easy to show treatment rules, we define a binary tree as our definition of interpretability. The interpretable treatment rules are generated as follows: 

1) Define a model family deemed as interpretable. We constrain a simple model to a binary tree with the maximum depth $d_{max}$ and the minimum number of observations $n_{obs}$ in each node. The model is denoted as $$r(\boldsymbol{X})=\sum_{m=1}^{M}w_{m}\mathds{1}\{\boldsymbol{X} \in R_{m}\} , s.t. \  d<=d_{max}, \abs{R_{m}}>n_{obs}.$$ The decision tree has two components $\{R,W\}$. $R=\{R_{m}\}, m=1,...,M$ is the covariate space partitioned by the splitting variables and the associated splitting rules, and $W=\{w_{m}\}$ is the predicted probability of $\widehat{d^{*}}(\boldsymbol{X})=1$ in the region $R_{m}$. The prediction of $p(\widehat{d^{*}}(\boldsymbol{X}_{i})=1)$ for units that end up in the region $R_{m}$ is $w_{m}$. 

2) Derive the best model $\widehat{r}$ by maximizing the following objective function: 
\begin{equation}
\label{equation: objective functions for tree F2'}
        \widehat{r}= \argmax_{\mathcal{W},\mathcal{R}} J(\widehat{r}) =  \argmax_{\mathcal{W},\mathcal{R}} \frac{1}{n}\sum_{m=1}^{M}\sum_{i \in R_{m}}\tilde{p}(\boldsymbol{X}_{i})\log w_{m}+(1-\tilde{p}(\boldsymbol{X}_{i}))\log (1-w_{m}))
\end{equation}
where $\tilde{p}(\boldsymbol{X}_{i})=p(\widehat{d^{*}}(\boldsymbol{X}_{i})=1)$ estimated from the MC samples from the posterior of $\widehat{f}$. The optimization of the function consists of two parts: determining covariate space $\{R_{m}\}$ and estimating the parameters $\{w_{m}\}$ in each space. First, we demonstrate how to estimate $w_{m}$ for the associated region $R_{m}$. $w_{m}$ is estimated by maximizing the objective function of the associated region $J(\widehat{r};R_{m})= \sum_{i \in R_{m}}\tilde{p}(\boldsymbol{x}_{i})\log w_{m}+(1-\tilde{p}(\boldsymbol{x}_{i}))\log (1-w_{m})$. The parameter $w_{m}$ can be derived by setting the derivative of $J(\widehat{r};R_{m})$ with respect to $w_{m}$ to $0$, leading to $w_{m}=\frac{1}{\abs{R_{m}}}\sum_{\boldsymbol{x}_{i} \in R_{m}}\tilde{p}(\boldsymbol{x}_{i})$. Second, we should find the optimal partitions of covariate space so that $J(\widehat{r})$ is maximized. Because of the computational infeasibility of choosing the best overall partition \cite{hastie2009elements}, we adopt a recursive binary splitting that starts from the top and proceeds with a greedy algorithm to determine a single best split pair $(X_{j},s)$ where $s$ is the cut-off value of variable $X_{j}$\cite{breiman1984classification}. The splitting criterias are provided in equations \ref{equation: splitting rules} as follows: 
\begin{equation}
\label{equation: splitting rules}
\begin{split}
&R_{j, s, l}=\left\{\boldsymbol{x} \mid x_{j} \leq s\right\} \\
&R_{j, s, r}=\left\{\boldsymbol{x} \mid x_{j}>s\right\}\\      
&R_{m} = R_{j,s,l}\cup R_{j,s,r} \\
&(j,s^{*}) = \arg \max_{s} J(R_{j,s,l})+J(R_{j,s,r})-J(R_{m})\\
&(j^{*},s^{*}) = \arg \max_{j} (j,s^{*}) \\
\end{split}
\end{equation}
The best partition $(X_{j^{*}},s^{*})$ is determined. Lastly, if one of the model constraints is met (i.e., $d>d_{max}$ or $\abs{R_{m}} <=n_{obs}$), the tree stops growing. The treatment decisions of individuals in leaf node $R_{m}$ are $\widehat{d^{*}}_{s}(\boldsymbol{X}_{i})=\mathds{1}\{w_{m}>0.5\}, \boldsymbol{X}_{i} \in R_{m}$. If two leaf nodes from the same branch have the same treatment decisions, these two leaf nodes will be merged and the treatment decision is determined based on the parameter of the merged region. 
\clearpage

\section{Deriving interpretable treatment rules using a logistic regression}
\label{subsec: simplify using logistic regression}
The logistic regression can uncover the relation of variables and the likelihood of $\widehat{d^{*}}(\boldsymbol{X})=1$. In some fields, logistic regression is commonly used to provide evidence regarding the optimal treatment for a (sub-)population. In this section, we demonstrate how to use logistic regression to simplify the optimal treatment rules. The procedure is based on the following two steps:  

1) Define a model family $r$ deemed as interpretable. We define a simple model family as a logistic regression with a basis fuction $\phi(\boldsymbol{X}) =(\phi_{1}(\boldsymbol{X}),..., \phi_{k}(\boldsymbol{X})) \in \mathbb{R}^{k}$, denoted as $$r(\boldsymbol{X})=\frac{1}{1+\text{exp}^{-\boldsymbol{w}\phi(\boldsymbol{X})}}, \boldsymbol{w} \in \mathbb{R}^{k},$$ where $\phi(\boldsymbol{X}_{i})$ could include non-linear relationships between variable $X_{j}, j \in \{1,...,p\}$, for instance, $\phi(\boldsymbol{X}_{i})=(1,X_{i1},X_{i2},X_{i1}^{2},X_{i1}*X_{i2})$;

2) Derive the best model $\widehat{r}$ from the model family. $\widehat{r}$ is obtained by maximizing the following objective function: 
    \begin{equation}
    \label{equation: objective function for logistic regression F2'}
       \widehat{r}= \argmax_{\mathcal{W}}J(r) = \argmax_{\mathcal{W}} \frac{1}{n}\sum_{i=1}^{n}\tilde{p}(\boldsymbol{X}_{i})\log r(\boldsymbol{X}_{i};\boldsymbol{w})+(1-\tilde{p}(\boldsymbol{X}_{i}))\log (1-r(\boldsymbol{X}_{i};\boldsymbol{w})).
    \end{equation}
where $\text{logit}(\widehat{r}(\boldsymbol{X}_{i}))=\widehat{\boldsymbol{w}}\phi(\boldsymbol{X}_{i})$. We use stochastic gradient descent to maximize $J(r)$ \cite{tsuruoka2009stochastic}. Parameters $\widehat{\boldsymbol{w}}$ are obtained according to $\widehat{w}_{j}^{t+1}=\widehat{w}_{j}^{t}+\eta\nabla_{w_{j}}J(r), j=1,..., k$, where $\nabla_{w_{j}}J(r)=\frac{\partial J(r)}{\partial w_{j}}=\left(\frac{1}{1+exp^{-\hat{\boldsymbol{w}}^{t}\phi(\boldsymbol{X}_{i})}}-\tilde{p}(\boldsymbol{X}_{i})\right)\phi_{j}(\boldsymbol{X}_{i})$, $\eta$ is the learning rate, $t$ is the number of iterations. Note that the gradient with respect to a single weight $w_{j}$ represents a very intuitive value: the difference between $\tilde{p}(\boldsymbol{X}_{i})$ and our estimate $\widehat{r}(\boldsymbol{X}_{i})$ for that observation, multiplied by the corresponding input value $\phi_{j}(\boldsymbol{X}_{i})$.

\clearpage

\appendix{Appendix E: Loss function parameterization}
\label{app: loss function intepretation}

Since the intuitive meaning of loss values is not clear when making decisions at different levels, we propose an additive form of the loss function. The parameterization of the loss function is shown in Table \ref{tab: additive loss function in our study},

\begin{table}[h]
    \centering
    \caption{additive form of loss function $L(Y(1),Y(0),t)$} 
    \begin{tabular}{c c c}
        \hline
         & $T=1$ & $T=0$ \\
        \hline
        $Y(1)=0,Y(0)=0$ & $c_{d}+c_{t}$ & $c_{d}$\\
        $Y(1)=0,Y(0)=1$ & $c_{d}+c_{t}$ & $0$\\
        $Y(1)=1,Y(0)=0$ & $c_{t}$ & $c_{d}$ \\
        $Y(1)=1,Y(0)=1$ & $c_{t}$ & $0$ \\
        \hline
    \end{tabular}
    \label{tab: additive loss function in our study}
\end{table} 
In this parameterization, $c_{t}$ denotes the loss caused by the treatment $T=1$ (e.g., the toxicity of the treatment), and $c_{d}$ denotes the loss caused by the undesirable outcome $Y=0$ (e.g., death). Under this loss parameterization, the optimal treatment rule is derived as follows:

\begin{equation} \label{equ: app proof optimal decision rule}
\begin{split}
\widehat{d^{*}}(\boldsymbol{X}_{i}) = & \mathds{1}\{\Delta L_{i} <0  \} \\
\Delta L_{i} = & \sum_{(j,k) \in \{0,1\}^{2}} (l_{jk}^{1}- l_{jk}^{0})\hat{\theta}_{ijk}\\
             = & c_{t}\widehat{\theta}_{i00}+(c_{d}+c_{t})\hat{\theta}_{i01} + (c_{t}-c_{d})\hat{\theta}_{i10}+  c_{t}\hat{\theta}_{i11} \\
             = & c_{t}+c_{d}(\widehat{\theta}_{i01}-\widehat{\theta}_{i10}) \\
        	     = & c_{t}+c_{d}(\widehat{\theta}_{i.1}-\widehat{\theta}_{i11} - (\hat{\theta}_{i1.}-\hat{\theta}_{i11}))\\
             = & c_{t}+c_{d}(\widehat{\theta}_{i.1}-\widehat{\theta}_{i1.}) \\
             = & c_{t} - c_{d}(\widehat{\theta}_{i1.}-\widehat{\theta}_{i.1}) \\
             =& c_{t} - c_{d}\widehat{\tau}(\boldsymbol{X}_{i})
\end{split}
\end{equation}
Then the optimal treatment decision is derived based on $\widehat{d^{*}}(\boldsymbol{X}_{i}) =  \mathds{1}\{\hat{\tau}(\boldsymbol{X}_{i})> \frac{c_{t}}{c_{d}}\}$, where $\frac{c_{t}}{c_{d}}$ can be regarded as a decision threshold. Individual $\boldsymbol{X}_{i}$ receives the treatment if the estimate of his/her treatment effect $\widehat{\tau}(\boldsymbol{X}_{i})$ is larger than the decision threshold, otherwise, control. The uncertainty of $\widehat{d^{*}}(\boldsymbol{X}_{i})$ is then estimated as $p(\widehat{d^{*}}(\boldsymbol{X}_{i})=1) \approx \frac{1}{D}\sum_{d}\mathds{1}\{\widehat{f}^{(d)}(\boldsymbol{X}_{i},1) - \widehat{f}^{(d)}(\boldsymbol{X}_{i},0) > \frac{c_{t}}{c_{d}}\}$. It is clear to see that under this loss function parameterization, the optimal treatment decision is independent of $\rho$ \cite{klausch2018estimating}. See discussion for more transformation of the inequality and practical meanings of $c_{t}$ and $c_{d}$ at different decision levels. 
\clearpage

\appendix{Appendix F: Summary statistics of the study population}
\label{Appendix F: Summary statistics of the study population}
\begin{center}
\begin{longtable}{l l l}
\caption{Summary statistics of the study population (n=27,057)}\\
\hline
\textbf{Variable} & \textbf{Overall$^{1}$} & \textbf{Chemo(\%)$^{2}$} \\
\hline
\endfirsthead
\multicolumn{3}{c}%
{\tablename\ \thetable\ -- \textit{Continued from previous page}} \\
\hline
\textbf{Variable} & \textbf{Overall$^{1}$} & \textbf{ACT(\%)$^{2}$} \\
\hline
\endhead
\hline \multicolumn{3}{r}{\textit{Continued on next page}} \\
\endfoot
\hline
\endlastfoot
Cancer diagnosis year              &                       &           \\
\ \ 2006-2008                          & 4,902 (18\%)          & 24        \\
\ \ 2009-2011                          & 5,726 (21\%)          & 34        \\
\ \ 2012-2014                          & 6,443 (24\%)          & 40        \\
\ \ 2015-2016                          & 9,986 (37\%)          & 38        \\
Age at diagnosis                   &                       &           \\
\ \ Mean (SD)                          & 70 (11)               &           \\
\ \ Median (IQR)                       & 71 (63 - 78)          &           \\
\ \ \textless{}=50                     & 1,410 (5.2\%)         & 59        \\
\ \ (50,60{]}                          & 3,467 (13\%)          & 54        \\
\ \ (60,70{]}                          & 8,578 (32\%)          & 47        \\
\ \ (70,80{]}                          & 8,916 (33\%)          & 28        \\
\ \ (80,90{]}                          & 4,419 (16\%)          & 5         \\
\ \ (90,102{]}                         & 267 (1.0\%)           & 0         \\
Gender                             &                       &           \\
\ \ FEMALE                             & 11,922 (44\%)         & 34        \\
\ \ MALE                               & 15,135 (56\%)         & 36        \\
Location                           &                       &           \\
\ \ Left-sided                         & 19,077 (71\%)         & 37        \\
\ \ Right-sided                        & 7,980 (29\%)          & 31        \\
Lymphatic invasion                 &                       &           \\
\ \ Absent                             & 7,501 (28\%)          & 32        \\
\ \ Present                            & 2,113 (7.8\%)         & 56        \\
\ \ Undetermined                       & 314 (1.2\%)           & 40        \\
\ \ unknown                            & 17,129 (63\%)         & 34        \\
The Number of lymph nodes examined &                       &           \\
\ \ Mean (SD)                          & 18 (11)               &           \\
\ \ Median (IQR)                       & 16 (11 - 22)          &           \\
\ \ \textless{}= 12                    & 8860 (32.7\%)         & 31        \\
\ \ \textgreater{}12                   & 18197 (67.3\%)        & 37        \\
MS                                 &                       &           \\
\ \ MSI                                & 695 (2.6\%)           & 30        \\
\ \ MSS                                & 3,043 (11\%)          & 55        \\
\ \ unknown                            & 23,319 (86\%)         & 33        \\
pT                                 &                       &           \\
\ \ T1                                 & 470 (1.7\%)           & 64        \\
\ \ T2                                 & 1,376 (5.1\%)         & 58        \\
\ \ T3                                 & 20,450 (76\%)         & 30        \\
\ \ T4                                 & 4,761 (18\%)          & 48        \\
pN                                 &                       &           \\
\ \ N0                                 & 13,144 (49\%)         & 9         \\
\ \ N1                                 & 9,163 (34\%)          & 57        \\
\ \ N2                                 & 4,750 (18\%)          & 65        \\
Grade$^{3}$                             &                       &           \\
\ \ Low grade                          & 21,723 (80\%)         & 35        \\
\ \ High grade                         & 3,175 (12\%)          & 40        \\
\ \ Unknown                            & 2,159 (8.0\%)         & 32        \\
Extramural angio invasion          &                       &           \\
\ \ Absent                             & 9,829 (36\%)          & 37        \\
\ \ Present                            & 337 (1.2\%)           & 53        \\
\ \ Undetermined                       & 521 (1.9\%)           & 42        \\
\ \ unknown                            & 16,370 (61\%)         & 33        \\
Intramural angio invasion          &                       &           \\ 
\ \ Absent                             & 8,440 (31\%)          & 34        \\
\ \ Present                            & 1,726 (6.4\%)         & 54        \\
\ \ Undetermined                       & 521 (1.9\%)           & 42        \\
\ \ unknown                            & 16,370 (61\%)         & 33        \\
Colon perforation                  &                       &           \\
\ \ Present                            & 517 (1.9\%)           & 41        \\
\ \ unknown                            & 26,540 (98\%)         & 35        \\
ASA performance status             &                       &           \\
\ \ 1                                  & 1,435 (5.3\%)         & 54        \\
\ \ 2                                  & 5,478 (20\%)          & 40        \\
\ \ 3                                  & 2,108 (7.8\%)         & 22        \\
\ \ 4                                  & 131 (0.5\%)           & 8         \\
\ \ unknown                            & 17,905 (66\%)         & 34        \\
The number of other tumors$^{4}$        &                       &           \\
\ \ 0                                  & 26,490 (98\%)         & 35        \\
\ \ 1                                  & 223 (0.8\%)           & 28        \\
\ \ 2                                  & 319 (1.2\%)           & 30        \\
\ \ 3                                  & 25 (\textless{}0.1\%) & 8         \\
ACT                                &                       &           \\
\ \ No                                 & 17,579 (65\%)         &           \\
\ \ FU with oxaliplatin                & 3,848 (14\%)          &           \\
\ \ FU without oxaliplatin             & 363 (1.3\%)           &           \\
\ \ Unknown$^{5}$                           & 5,267 (19\%)          &          
\end{longtable}
\label{tab: summary statistics of the study population}
$^{1}$n (\%); $^{2}$proportion of patients receiving ACT; $^{3}$Low-grade histology includes moderate to well-differentiated. High-grade histology includes poorly differentiated and undifferentiated; $^{4}$the number of non-primary C18 tumors diagnosed simultaneously; $^{5}$ACT may include oxaliplatin but we don’t have available information to identify specific regimen \\
\end{center}

\clearpage

\appendix{Appendix G: Additional results}
\label{Appendix G: Results of accuracy, precision, and recall of three treatment rules}

\begin{figure}[H]
    \centering
    \includegraphics[width=1\textwidth]{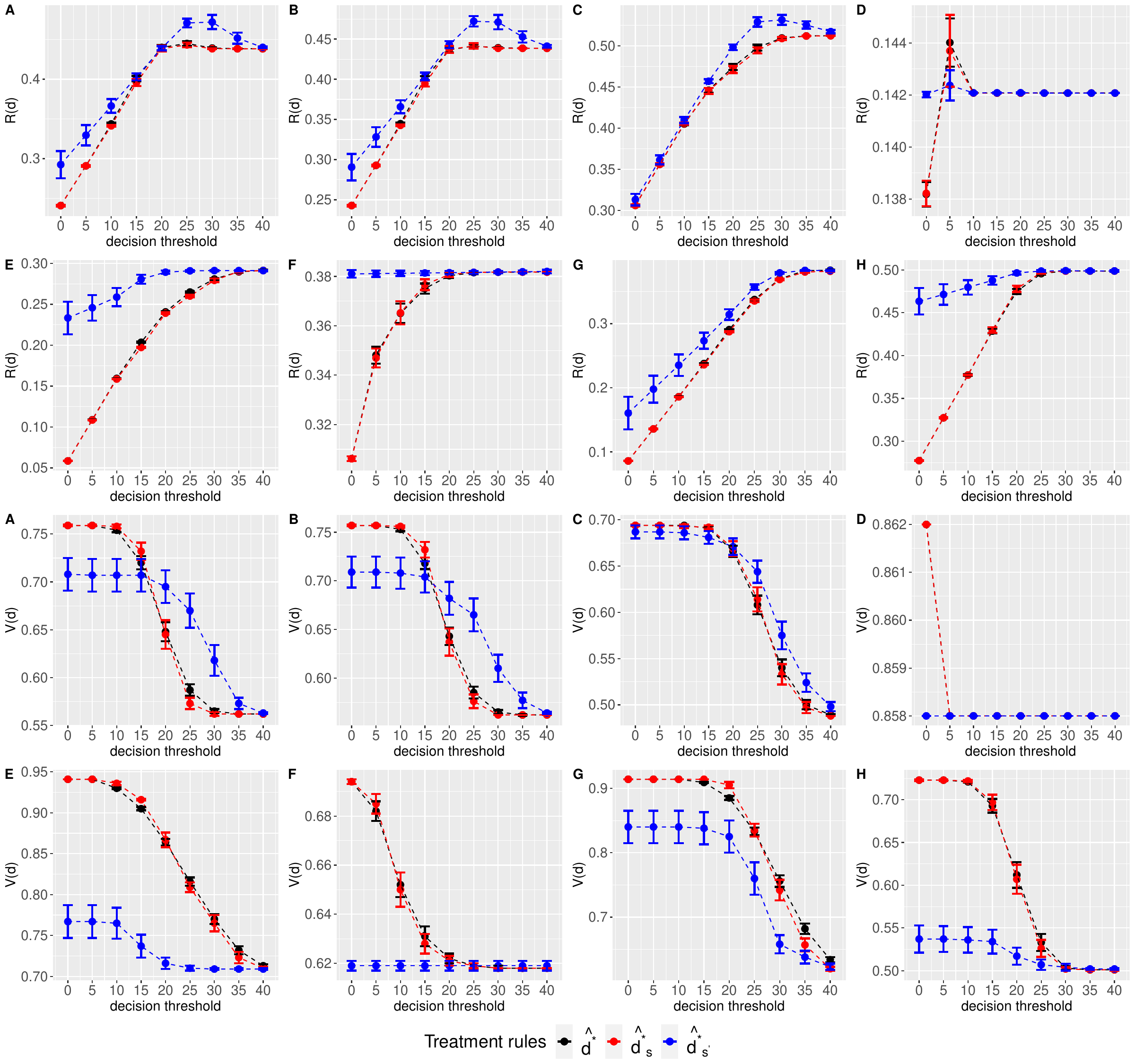}
    \caption{Average loss $R$ and average outcome $V$ of  $\widehat{d^{*}}(\boldsymbol{X}_{i})$, $\widehat{d^{*}}_{s}(\boldsymbol{X}_{i})$, and $\widehat{d^{*}}_{s'}(\boldsymbol{X}_{i})$, where $f(\boldsymbol{X},T)$ for estimating  $\widehat{d^{*}}_{s'}(\boldsymbol{X}_{i})$ and $r(\boldsymbol{X})$ for estimating $\widehat{d^{*}}_{s}(\boldsymbol{X}_{i})$ are constructed using binary trees with $d_{max}=2$, $n_{obs}=5$, the decision threhold denotes the ratio of the loss due to the treatment to the loss due to the undesirable outcome. }
    \label{fig: simulation_loss_outcome_TREE_depth1} 
\end{figure}

\begin{figure}[H]
    \centering
    \includegraphics[width=0.95\textwidth]{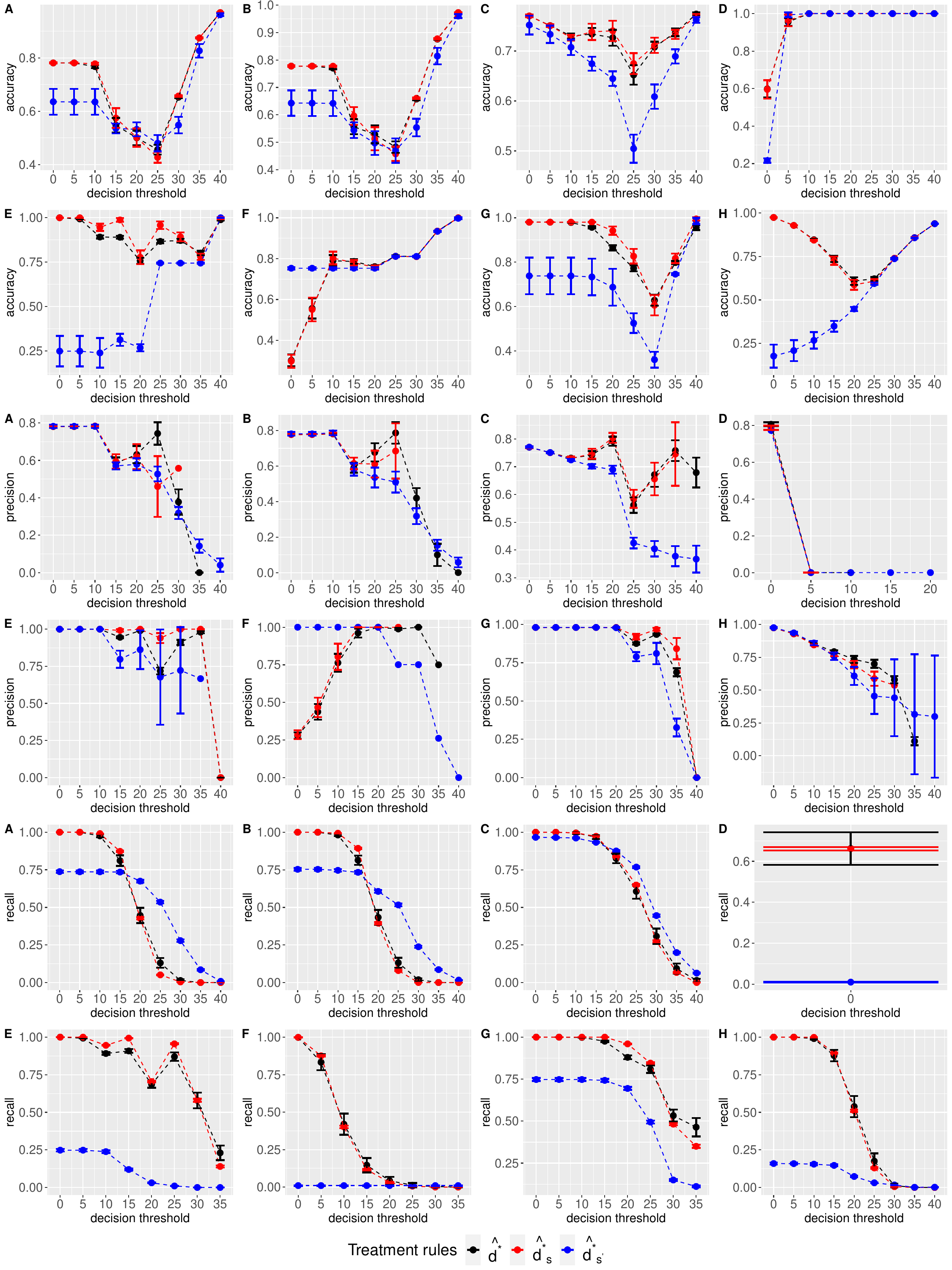}
    \caption{Accuracy, precision, and recall of three treatment rules $\widehat{d^{*}}(\boldsymbol{X}_{i})$, $\widehat{d^{*}}_{s}(\boldsymbol{X}_{i})$, and $\widehat{d^{*}}_{s'}(\boldsymbol{X}_{i})$, where $f(\boldsymbol{X},T)$ for estimating  $\widehat{d^{*}}_{s'}(\boldsymbol{X}_{i})$ and $r(\boldsymbol{X})$ for estimating $\widehat{d^{*}}_{s}(\boldsymbol{X}_{i})$ are constructed using binary trees with $d_{max}=2$, $n_{obs}=5$. It's worth noting that as the decision threshold increases, the count of individuals for whom the predicted optimal treatment or the true optimal treatment are 1 decreases. The denominator in precision or recall may become zero, which is not reflected in the plot.}
    \label{fig: simulation_BART_TREE1_acc_precision_recall} 
\end{figure}

\begin{figure}[H]
    \centering
    \includegraphics[width=0.95\textwidth]{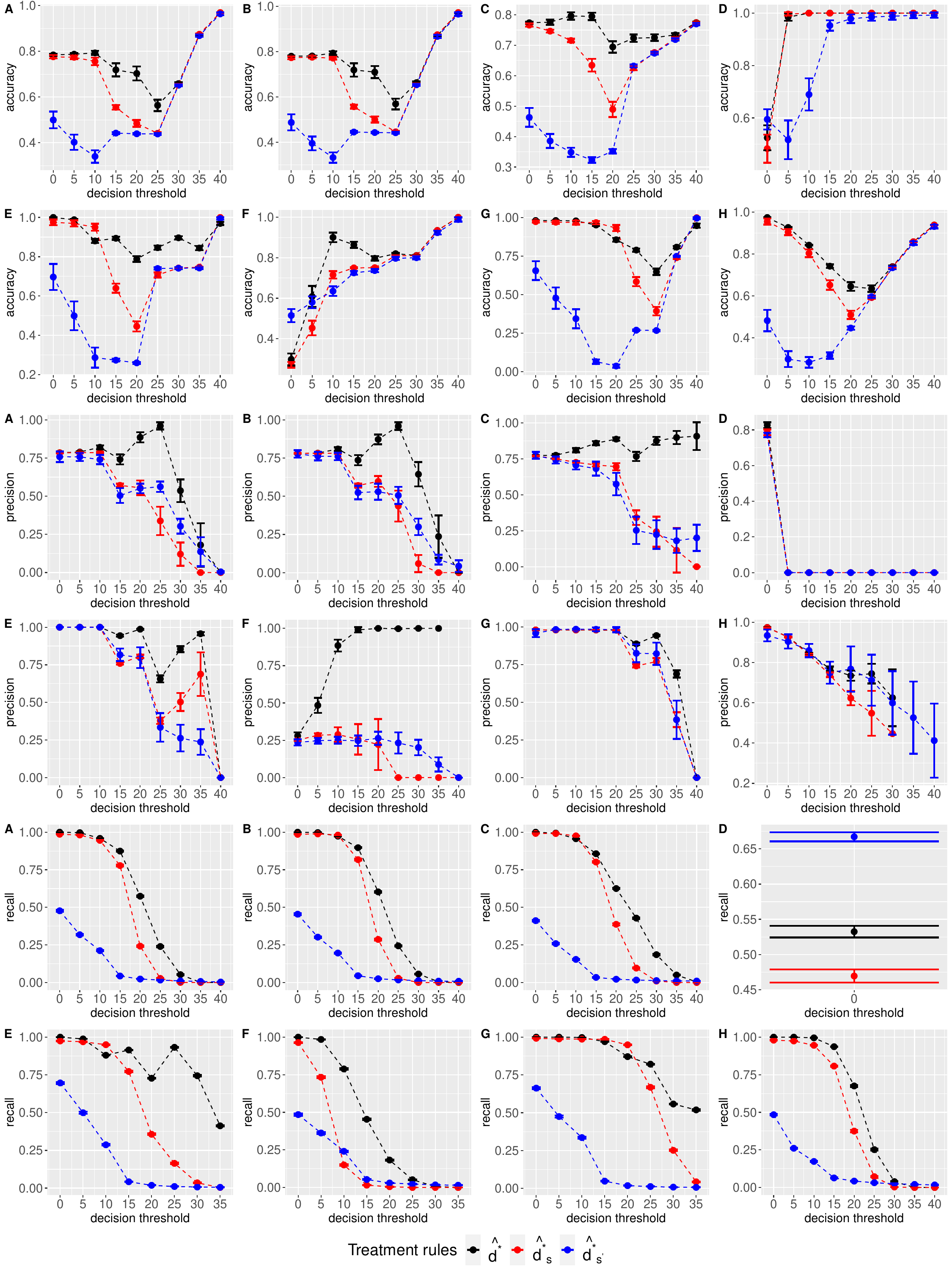}
    \caption{Accuracy, precision, and recall of three treatment rules $\widehat{d^{*}}(\boldsymbol{X}_{i})$, $\widehat{d^{*}}_{s}(\boldsymbol{X}_{i})$, and $\widehat{d^{*}}_{s'}(\boldsymbol{X}_{i})$, where $f(\boldsymbol{X},T)$ for estimating  $\widehat{d^{*}}_{s'}(\boldsymbol{X}_{i})$ and $r(\boldsymbol{X})$ for estimating $\widehat{d^{*}}_{s}(\boldsymbol{X}_{i})$ are logistic regression. It's worth noting that as the decision threshold increases, the count of individuals for whom the predicted optimal treatment or the true optimal treatment are 1 decreases. The denominator in precision or recall may become zero, which is not reflected in the plot.}
    \label{fig: simulation_BART_LReg_acc_precision_recall} 
\end{figure}

\end{document}